\documentclass{elsart}

\usepackage{graphicx}
\usepackage{amssymb}

\begin{document}

\begin{frontmatter}

\title{Four-dimensional Yang-Mills theory\\with a three-dimensional fermion membrane}

\author{Arata~Yamamoto}
\ead{a-yamamoto@ruby.scphys.kyoto-u.ac.jp}
\address{Department of Physics, Faculty of Science, Kyoto University, \\Kitashirakawa, Sakyo, Kyoto 606-8502, Japan}

\begin{abstract}
We study the four-dimensional Yang-Mills theory in the presence of a three-dimensional membrane of fermions by lattice Monte Carlo simulations.
We analyze the phase structure of this theory at finite temperature.
Below the phase transition temperature of the pure Yang-Mills theory, we obtain an unconventional phase with spatially-nonuniform vacuum.
In this phase, the expectation value of the Polyakov loop is finite on the membrane, and it exponentially decays to zero outside the membrane.
\end{abstract}

\end{frontmatter}

\section{Introduction}

As a theoretical interest in the gauge field theory, we can consider the theory which contains the gauge field and the matter field in different dimensions.
Such a theory often shows characteristic properties \cite{Kaplan:1992bt,Gorbar:2001qt,Kaplan:2009kr,Narayanan:2009ag}.
In condensed matter physics, it has been realized in trapped electron systems, such as graphene \cite{CastroNeto:2009zz}.
Recently, graphene is investigated in lattice gauge theory \cite{Hands:2008id,Drut:2008rg,Armour:2009vj,Araki:2010gj}.
In the gauge/string duality, this kind of setup has been frequently used in terms of D-brane \cite{Aharony:1999ti}.

\begin{figure}[t]
\begin{center}
\includegraphics[scale=0.5]{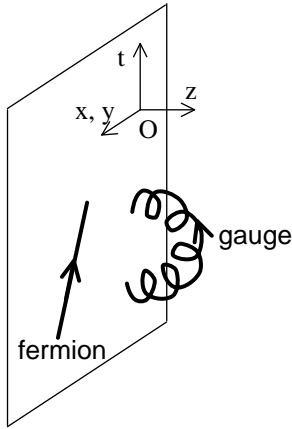}
\caption{\label{fig1}
The schematic figure of the (3+1)-dimensional gauge theory with (2+1)-dimensional fermions.
The fermions are localized at $z=0$.
}
\end{center}
\end{figure}

Motivated by these works, we consider the Yang-Mills theory coupled with fermions.
We depict our theoretical setup schematically in Fig.~\ref{fig1}.
The Yang-Mills field lives in the (3+1)-dimensional space-time, while the fermion field lives only on the (2+1)-dimensional hyperplane at $z=0$.
The existence of this fermion membrane breaks the Lorentz invariance and the translational invariance in the $z$-direction.
From the viewpoint of the gauge field, the fermion field is localized in the low-dimensional subsystem.
From the viewpoint of the fermion field, the background gauge field has one extra dimension.

This theory is renormalizable.
Counter terms are introduced for the gauge field in four dimension and for the fermion field in three dimension at $z=0$.
There is no additional ultraviolet divergence, unlike two-dimensional fermions in the four-dimensional Yang-Mills theory \cite{Gorbar:2001qt,Narayanan:2009ag}.
This theory is classically conformal when the fermions are massless.
There is no dimensional parameter except for the bare fermion mass, unlike the three-dimensional Yang-Mills theory.
The dynamics is uniquely determined through dimensional transmutation.

In this Letter, we study this system by means of lattice Monte Carlo simulations.
We focus on how the fermion membrane affects the vacuum structure at finite temperature.
In the finite-temperature formalism, the $t$-direction is compactified in the Euclidean metric with a periodic boundary condition for the gauge field and with an antiperiodic boundary condition for the fermion field.

\section{Formalism}
This system is formulated by the standard technique of lattice gauge theory.
For the gauge part of the lattice action, we used the Wilson gauge action
\begin{eqnarray}
S_G = \beta \sum_{x,\mu,\nu} \left[ 1 - \frac{1}{N_c} {\rm ReTr} U_{\mu \nu} (x) \right],
\end{eqnarray}
where $U_{\mu \nu} (x)$ is the plaquette variable, i.e., the minimal gauge invariant loop.
The lattice spacing $a$ is determined by the dimensionless parameter $\beta=2N_c/g^2$.
Because this system is equivalent to the pure gauge theory in large $z$, we can set the same physical scale as in the pure gauge theory. 
For the fermion part, we used the three-dimensional staggered fermion,
\begin{eqnarray}
S_F &=& \frac{1}{2} a^2 \sum_{x,i} \eta_i (x) \Bigl[ \bar{\chi}(x) U_i(x) \chi(x+\hat{i})\nonumber \\
&&  - \bar{\chi}(x) U^\dagger_i(x-\hat{i}) \chi(x-\hat{i}) \Bigr] + a^3 m \sum_{x} \bar{\chi}(x) \chi(x) ,
\end{eqnarray}
where $\chi(x)$ is the spinorless staggered field and $\eta_i (x)$ is the staggered phase \cite{Burden:1986by}.
The summations are taken over the three-dimensional plane at $z=0$.
In the continuum limit, this fermion action corresponds to the standard three-dimensional form $\sum \bar{\psi} (\sigma_i D_i \pm m)\psi$ and the fermion field is the four-flavor two-component spinor.
The four flavors contain two positive-mass states of $+m$ and two negative-mass states of $-m$.
This fermion field is equivalently rewritten as the two-flavor four-component spinor under a $4\times 4$ representation of the Dirac matrices \cite{Burden:1986by,Pisarski:1984dj}.
This four-component spinor construction preserves parity and time-reversal invariances.

To generate the gauge configurations under the full lattice action $S_G+S_F$, we made use of the Hybrid Monte Carlo algorithm.
The gauge configuration with the lattice volume $N_s^3\times N_t$ includes the dynamical fermions on the three-dimensional volume $N_s^2\times N_t$.
The parameters are set at $N_c=3$ and $\beta= 5.7$, and the corresponding lattice spacing is $a\simeq 0.19$ fm.
The bare fermion mass is set at $ma=0.2$.
To simulate several values of temperature, we changed the temporal extent $N_t$ with the fixed lattice spacing.
The physical temperature is given as $T = 1/(N_ta)$ with $a^{-1}\simeq 1$ GeV.
The lattice sizes of the gauge configurations are listed in Table \ref{tab1}.

\section{Results}
\begin{figure}[t]
\begin{center}
\includegraphics[scale=1.2]{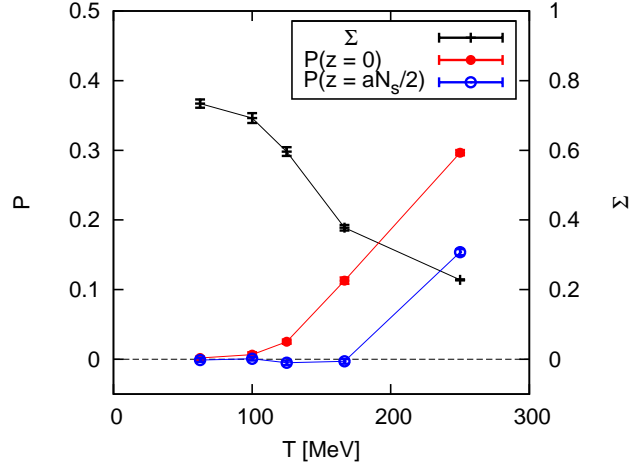}
\caption{\label{fig2}
The Polyakov loop $P$ and the fermion condensate $\Sigma$ with $N_s=16$.
}
\end{center}
\end{figure}

First, we measured the fermion condensate
\begin{equation}
\Sigma= -a^2\langle \bar{\psi} \psi \rangle
\end{equation}
 at the $z=0$ plane.
We show the numerical data in Table \ref{tab1} and Fig.~\ref{fig2}.
Note that the fermion condensate is always finite because of the finite fermion mass.
The fermion condensate decreases as temperature increases, and its derivative seems to be large in $125 \ {\rm MeV} < T < 167$ MeV.
This suggests that the remnant of spontaneous symmetry breaking contributes to the fermion condensate in low temperature and it is restored in high temperature.
The broken symmetry is flavor symmetry, not chiral symmetry, because there is no chiral symmetry in three dimension.
In the massless and continuum limit, the symmetry breaking pattern is considered to be $U(N_f) \to U(N_f/2) \times U(N_f/2)$, as three-dimensional QCD \cite{Verbaarschot:1994ip,Damgaard:1998yv}.

Next, we calculated the expectation value of the Polyakov loop
\begin{equation}
P = \left\langle \frac{1}{N_c} {\rm Tr} \prod_t U_0(\vec{x}, t) \right\rangle,
\end{equation}
which is a good indicator for confinement.
Unlike the fermion condensate, we can define the Polyakov loop not only on the fermion membrane ($z=0$) but also outside it ($z\ne 0$).
We show the Polyakov loop value at $z=0$ and $z=aN_s/2$ in Table \ref{tab1} and Fig.~\ref{fig2}.
In large $z$ limit, this theory should be equivalent to the pure Yang-Mills theory.
The SU(3) pure gauge theory has the first-order phase transition at $T_c\simeq 250$ MeV, e.g., $\beta_c\simeq 5.69$ with $N_t=4$ \cite{Iwasaki:1992ik}
The data of $z=aN_s/2$ is consistent with this expectation.
On the other hand, the Polyakov loop at $z=0$ shows a different behavior.
It is finite even at $T\simeq 125$ MeV.
As a result, we obtain an unconventional phase in $125 \ {\rm MeV} \le T < 250$ MeV.
In this phase, the Polyakov loop value is nonzero at $z=0$, whereas it is almost zero in large $z$.
This phase is interpreted as confinement phase with partial deconfinement around the fermion membrane.

\begin{table}[t]
\begin{center}
\renewcommand{\tabcolsep}{0.5pc} 
\renewcommand{\arraystretch}{1} 
\caption{\label{tab1}
Numerical data of simulations.
The temporal extent $N_t$, the spatial extent $N_s$, the physical temperature $T$, the fermion condensate $\Sigma$ and the Polyakov loop $P$ are listed.
The statistical errors are shown in parentheses.
}
\begin{tabular}{cccccc}
\hline
$N_t$ & $N_s$ & $T$ [MeV] & $\Sigma$ & $P(z=0)$ & $P(z=aN_s/2)$\\
\hline
16 & 16 &  63 & 0.734(12) & 0.0017(22) & -0.0011(28) \\
10 & 16 & 100 & 0.692(14) & 0.0065(34) &  0.0007(27) \\
 8 & 16 & 125 & 0.596(13) & 0.0252(31) & -0.0051(30) \\
 6 & 12 & 167 & 0.369(7) &  0.1025(44) &  0.0121(41) \\
 6 & 16 & 167 & 0.378(8) &  0.1128(46) & -0.0030(33) \\
 6 & 24 & 167 & 0.379(6) &  0.0999(30) & -0.0008(22) \\
 4 & 16 & 250 & 0.228(1) &  0.2966(38) &  0.1538(43) \\
\hline
\end{tabular}
\end{center}
\end{table}

The existence of this phase is understood as follows.
Apart from the membrane, the system has the first-order phase transition of the pure Yang-Mills theory at $T_c\simeq 250$ MeV.
On the membrane, however, the theory includes the dynamical fermions.
This makes the coupling constant smaller and the transition temperature lower locally around the membrane.
Therefore, the fermion membrane induces partial deconfinement below the transition temperature of the pure Yang-Mills theory.

\begin{figure}[t]
\begin{center}
\includegraphics[scale=1.2]{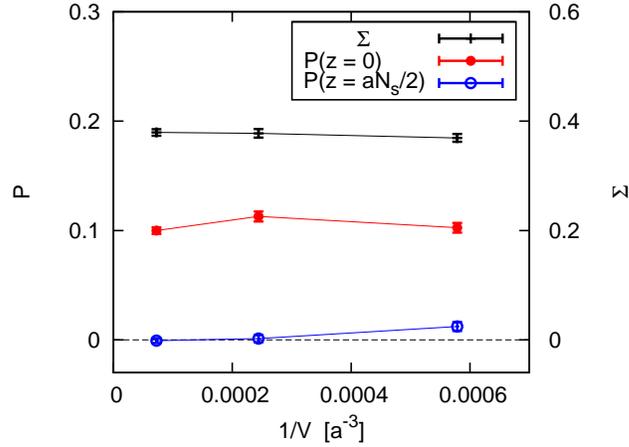}
\caption{\label{fig3}
The volume dependence of the Polyakov loop $P$ and the chiral condensate $\Sigma$ with $N_t=6$.
}
\end{center}
\end{figure}

In the present simulations, we cannot determine the temperature and the order of the phase transition.
These properties would depend on the fermion mass and the flavor number.
To validate the existence of this phase, we here check the dependence on the spatial volume $V=a^3N_s^3$ at fixed temperature.
In Fig.~\ref{fig3}, we plot the fermion condensate and the Polyakov loop at $T\simeq 167$ MeV as a function of the inverse volume $1/V$.
All these are insensitive to the volume.
In particular, the Polyakov loop at $z=0$ seems to remain finite in the infinite volume limit $1/V \to 0$.
Thus, at least, the calculation at $T\simeq 167$ MeV, i.e., $N_t=6$, lies in this phase.

\begin{figure}[t]
\begin{center}
\includegraphics[scale=1.2]{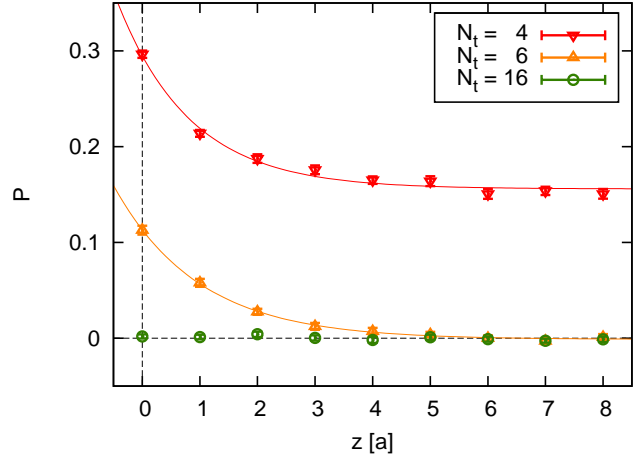}
\caption{\label{fig4}
The $z$-dependence of the Polyakov loop $P$ with $N_s=16$.
The curves are the best-fit functions of $C\exp(-z/z_0)+P_0$.
}
\end{center}
\end{figure}

In this phase, the vacuum structure is spatially-nonuniform in the $z$-direction.
In Fig.~\ref{fig4}, we show the Polyakov loop value as a function of $z$.
Since the distribution is symmetric about $z=0$, we only show the region of $0 \le z \le aN_s/2$.
The numerical data is well fitted by $C\exp(-z/z_0)+P_0$.
At $T\simeq 167$ MeV, i.e., $N_t=6$, the Polyakov loop value is nonzero at $z=0$ and decreases exponentially in $z>0$.
The best-fit damping parameter is $z_0 \simeq 0.3$ fm.
Taking into account the region of $z \le 0$, the thickness of the deconfinement layer is roughly 1 fm.

\begin{figure}[t]
\begin{center}
\includegraphics[scale=1.2]{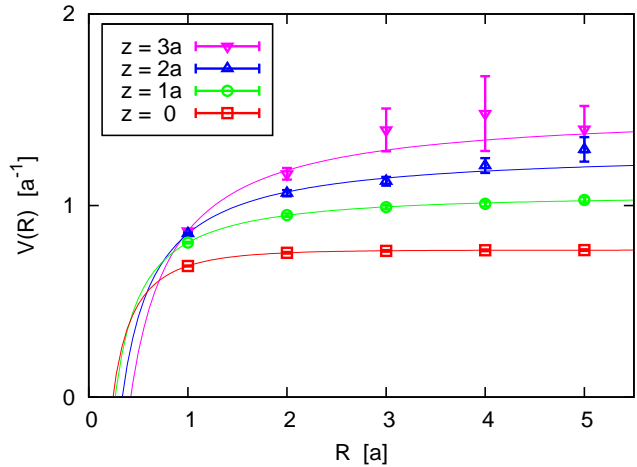}
\caption{\label{fig5}
The color averaged potential $V(R)$ with $N_s=16$ and $N_t=6$.
The curves are the best-fit functions of $-A\exp(-MR)/R+V_0$.
}
\end{center}
\end{figure}

As another observable, we consider the color averaged potential between static color sources.
The color averaged potential $V(R)$ is extracted from the correlation function of the Polyakov loop and the anti-Polyakov loop,
\begin{equation}
e^{-\frac{V(R)}{T}} = \left\langle \frac{1}{N_c^2} \biggl[{\rm Tr} \prod_{t_1} U_0(\vec{x}_1, t_1)\biggr] \biggl[{\rm Tr} \prod_{t_2} U_0^\dagger(\vec{x}_2, t_2)\biggr]\right\rangle
\end{equation}
with $R=|\vec{x}_1-\vec{x}_2|$.
In the case of $N_c=3$, the color averaged potential includes the color-singlet and color-octet components.
In a deconfinement medium, the potential is Debye screened as the color sources are separated.
We measured the color averaged potential from the Polyakov loop and the anti-Polyakov loop at the same $z$ plane.
In Fig~\ref{fig5}, we plot the resultant potential at $T\simeq 167$ MeV as for several values of $z$.
We also show the best-fit function of $-A\exp(-MR)/R+V_0$.
At $z=0$, the potential is strongly screened and the screening mass is $M\simeq 1$ GeV.
As $z$ increases, the screening mass decreases as $M\to 0$.
In this phase, we can change the potential from the Debye screened form to the Coulomb plus confining form by going away from the membrane, without changing temperature.

In summary, we discussed the four-dimensional Yang-Mills system with a three-dimensional fermion membrane, and found that the fermion membrane induces a deconfinement layer, which is about 1 fm thick, in confinement phase.
If we introduce fermions with the larger flavor number, we would obtain a conformal layer at zero temperature.
In general, we can patch not only a membrane but also various different vacuums in the same manner.
This kind of theoretical study is one possible approach for spatially-nonuniform vacuum in lattice gauge theory.

\section*{Acknowledgments}
The author is supported by a Grant-in-Aid for Scientific Research [(C) No.~20$\cdot$363].
This work was supported by the Global COE Program, ``The Next Generation of Physics, Spun from Universality and Emergence,'' at Kyoto University.
The lattice QCD simulations were carried out on NEC SX-8R in Osaka University.

\end{document}